\documentclass[
  reprint,
  nofootinbib,
  amsmath,amssymb,
  aps,
  floatfix,
]{revtex4-2}

\usepackage{xcolor}
\usepackage{graphicx}
\usepackage{slashed}
\usepackage[compat=1.1.0]{tikz-feynman}
\tikzfeynmanset{warn luatex=false}

\usepackage[hidelinks]{hyperref}
\usepackage{cleveref}

\crefname{figure}{Fig.}{Figs.}
\Crefname{figure}{Fig.}{Figs.}
\crefname{equation}{Eq.}{Eqs.}
\Crefname{equation}{Eq.}{Eqs.}
\crefname{section}{Section}{Sections}
\Crefname{section}{Section}{Sections}
\crefname{appendix}{Appendix}{Appendices}
\Crefname{appendix}{Appendix}{Appendices}

\newcommand{\nn}{\nonumber}
\newcommand{\cA}{\mathcal{A}}
\newcommand{\psibar}{\overline{\psi}}
\newcommand{\Cpauli}{C}

\newcommand{\SoftFactor}[1]{S^{(#1)}}
\newcommand{\SoftFactorLog}[1]{\SoftFactor{#1}_{\log}}
\newcommand{\SoftFactorFinite}[1]{\SoftFactor{#1}_{\mathrm{finite}}}
\newcommand{\SoftFactorFlavor}[2]{S^{(#1)#2}}

\newcommand{\QEDSoftLeading}{\SoftFactor{-1}}
\newcommand{\QEDSoftSubleading}{\SoftFactor{0}}
\newcommand{\QEDSoftSubleadingLog}{\SoftFactorLog{0}}
\newcommand{\QEDSoftSubleadingFinite}{\SoftFactorFinite{0}}
\newcommand{\QEDSoftSubleadingPauli}{\QEDSoftSubleading_{\mathrm{QED+P}}}

\newcommand{\NLSMSoftLeading}{\SoftFactor{0}}
\newcommand{\NLSMSoftSubleading}{\SoftFactor{1}}
\newcommand{\NLSMSoftSubleadingLog}{\SoftFactorLog{1}}
\newcommand{\NLSMSoftSubleadingFinite}{\SoftFactorFinite{1}}
\newcommand{\NLSMSoftSubleadingFlavor}[1]{\SoftFactorFlavor{1}{#1}}

\begin{document}

\preprint{APS/123-QED}
\hfill UUITP-11/26

\title{Running soft theorems in effective field theory}

\author{Jonah Berean-Dutcher}
\affiliation{%
 University of British Columbia, Vancouver, V6T 1Z1, Canada\\
 Institut des Hautes Études Scientifiques, 91440 Bures-sur-Yvette, France
}%

\author{Maria Derda}
\affiliation{%
 Department of Physics and Astronomy, Uppsala University, 75120 Uppsala, Sweden
}%

\author{Riley A. Stewart}
\affiliation{%
 University of British Columbia, Vancouver, V6T 1Z1, Canada
}%

\date{\today}

\begin{abstract}
We study soft theorems for photons and pions with local effective field theory (EFT) corrections up to one loop order. In massive QED with a dimension-five Pauli operator, the local analytic part of the subleading soft photon theorem is renormalized by the hard integration region. On the other hand, as was previously known, the logarithmic soft terms do not receive corrections from higher-dimensional operators. We find that analogous structure appears in the double-soft pion theorem: the coefficients of logarithmic contributions are controlled entirely by the Nonlinear Sigma Model (NLSM) Lagrangian, while the Wilson coefficients of four-derivative operators in Chiral EFT appear in local subleading terms, in a form that is dictated by their RG equations.
\end{abstract}

\maketitle

Soft theorems organize the infrared behavior of scattering amplitudes by relating processes with soft external particles to lower-point processes without the soft particles.
For EFTs involving photons or pions, these statements include the Low-Burnett-Kroll theorem for soft photons \cite{Weinberg:1965nx, Burnett:1967km}, the Adler zero for soft pions \cite{Adler:1964um}, and the double-soft pion theorem \cite{Weinberg:1966gjf,Arkani-Hamed:2008owk,Kampf:2013vha,Cachazo:2015ksa,Du:2015esa,Rodina:2021isd}. 
In these examples, the leading order soft behavior is dictated by the global symmetries in the infrared theory, while subleading terms can contain non-universal local contributions controlled by EFT operators \cite{Elvang:2016qvq}. In this paper, we study the interplay between the renormalization group (RG) evolution of such operators and the subleading soft theorems for photons and pions. As we will see, the non-universal subleading soft structures share the same qualitative features, while their detailed form is controlled by the RG equations for the Wilson coefficients of the EFT.

Firstly, we discuss the soft photon theorem in QED with massive fermions, deformed by a Pauli-type operator \cite{Engel:2021ccn,Krishna:2023fxg,Laddha:2018myi,Sahoo:2018lxl,Elvang:2016qvq,AtulBhatkar:2018kfi, Bern:2014vva}. At one loop, the soft theorem for a scattering amplitude with $n$ hard particles and a soft photon has a schematic form 
\begin{align}
\label{eq:soft_photon_schematic}
    \lim_{\tau\to 0}\cA_{n+\gamma}
    = \left(\frac{1}{\tau}S^{(0)} + \log\tau\, S^{(1)}_{\log} + S^{(1)}_{\mathrm{finite}}\right)\cA_{n}
    + \mathcal{O}(\tau)\,,
\end{align}
where the soft limit is taken by scaling the photon momentum as $q\to \tau q$ with $\tau\to 0$. Here, $S^{(0)}$ is the familiar leading soft photon factor \cite{Weinberg:1965nx,Low:1958sn, Burnett:1967km, Yennie:1961ad, Bern:2014vva, Engel:2023ifn, Engel:2023rxp, DelDuca:1990gz}. As was shown in \cite{Engel:2021ccn,Krishna:2023fxg,Laddha:2018myi,Sahoo:2018lxl,AtulBhatkar:2018kfi,Engel:2023ifn}, the subleading soft factor $S^{(1)}$ is dominated by infrared logarithms, denoted as $S^{(1)}_{\log}$. Those logarithmic terms do not depend on possible higher-dimensional operators in the EFT \cite{Sahoo:2018lxl, Krishna:2023fxg, AtulBhatkar:2018kfi}. The remaining term $S^{(1)}_{\mathrm{finite}}$ in the subleading soft factor includes the non-universal local terms controlled by higher-dimensional EFT operators \cite{Elvang:2016qvq}. In this paper, we compute one loop corrections to $S^{(1)}_{\mathrm{finite}}$ which renormalize those local terms.

Next, we consider chiral EFT describing the interactions of massless $SU(N)$ pions and compute the loop order corrections to the double soft pion theorem, which for the best of our knowledge have not appeared in the literature before. Analogously to soft photon theorem in \cref{eq:soft_photon_schematic}, we can organize the double soft theorem for an amplitude with $n+2$ pions as
\begin{align}
    \lim_{\tau\rightarrow 0} \cA_{n+2}
    = \left(S^{(0)} + \tau\log\tau\, S^{(1)}_{\log} + \tau\, S^{(1)}_{\mathrm{finite}}\right)\cA_{n}+ \mathcal{O}(\tau^2)
\end{align}
where the soft parameter $\tau$ is used to uniformly scale the magnitude of two external pion momenta to zero $q_a\to\tau q_a$ and $q_b\to\tau q_b$. As before, $S^{(0)}$ denotes the well-known leading double soft pion factor \cite{Weinberg:1966gjf}. At subleading order, loop level corrections yield the dominant $S^{(1)}_{\log}$ terms, which again are independent of the higher-derivative operators in the theory. The part of the soft factor $S^{(1)}_{\mathrm{finite}}$ carries non-universal dependence on the Wilson coefficients of the chiral EFT, which was first derived in \cite{Rodina:2021isd}. The one loop running of $S^{(1)}_{\mathrm{finite}}$ is controlled by the running of these Wilson coefficients \cite{Gasser:1983yg}.

In each of these two example EFTs the one loop, subleading soft theorem contains contributions that may be separated: (1) non-universal with dependence on Wilson coefficients of higher-dimensional operators, and (2) universal with dependence on only the lowest energy operators of the EFT. The scale dependence of the the non-universal terms is governed by the EFT.

Since soft theorems are a relation between amplitudes, the soft factors themselves must also be independent of the RG scale $\mu$,
\begin{align}
    \mu \frac{d}{d\mu}S^{(i)} = 0\,. \label{eq:rg-invariance-of-soft-factor}
\end{align}
After a choice of scheme, \cref{eq:rg-invariance-of-soft-factor} can be alternatively viewed as implying the corresponding RG equations for the participating Wilson coefficients. We will highlight how this is realized in each example EFT.

The outline of the paper is as follows. In \cref{sec:soft-photons}, we present the hard integration region corrections to the soft photon theorem at one loop, relegating technical details of the derivation -- and discussion of scheme dependent features -- to \cref{sec:appendix-qed-examples}. The one loop corrections to the double soft pion theorem are discussed in \cref{sec:chiraleft} for a specific ordering of flavor-ordered amplitudes. The full one loop double soft theorem for flavor dressed amplitudes is provided in \cref{sec:appendix-full-double-soft-pions}. Throughout, we use dimensional regularization in $d=4-2\varepsilon$ dimensions. In computational checks we evaluated one loop integrals using \textit{Package}-\texttt{X} \cite{Patel:2015tea,Patel:2016fam}. The soft expansion is taken after loop integration and expansion in the dimensional regulator \cite{Bern:2014oka, Broedel:2014bza, Bonocore:2014wua}.

\section{Soft photons}\label{sec:soft-photons}
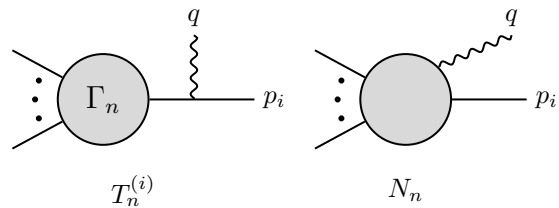
\begin{figure}[b]
    \centering
    \begin{tikzpicture}[thick, line width=0.75pt, baseline=(current bounding box.center)]
        \tikzset{
            solid blob/.style={circle, draw, fill=gray!30, minimum size=1.2cm, inner sep=0pt, font=\large},
            blob/.style={circle, draw, fill=black, minimum size=1.5pt, inner sep=0pt},
            label/.style={font=\normalsize}
        }

        \node[solid blob] (b1) at (0, 0) {\(\Gamma_n\)};
        \coordinate (tl1) at (-1.2, 0.65);
        \coordinate (bl1) at (-1.2, -0.65);
        \coordinate (v1) at (1.2, 0);
        \coordinate (tr1) at (2, 0);
        \coordinate (gamma1) at (1.2, 0.85);

        \node[blob] at (-0.85, 0.25) {};
        \node[blob] at (-0.9, 0) {};
        \node[blob] at (-0.85, -0.25) {};

        \draw (tl1) -- (b1);
        \draw (bl1) -- (b1);
        \draw (b1) -- (v1) -- (tr1) node[right, label] {$p_i$};
        \draw[/tikzfeynman/photon] (v1) -- (gamma1) node[above, label] {$q$};
        \node[label] at (0.4, -1.2) {\(T_n^{(i)}\)};

        \node[solid blob] (b2) at (4, 0) {};
        \coordinate (tl2) at (2.8, 0.65);
        \coordinate (bl2) at (2.8, -0.65);
        \coordinate (tr2) at (5.6, 0);
        \coordinate (gamma2) at (5.4, 0.85);

        \node[blob] at (3.15, 0.25) {};
        \node[blob] at (3.1, 0) {};
        \node[blob] at (3.15, -0.25) {};

        \draw (tl2) -- (b2);
        \draw (bl2) -- (b2);
        \draw (b2) -- (tr2) node[right, label] {$p_i$};
        \draw[/tikzfeynman/photon] (b2.45) -- (gamma2) node[above, label] {$q$};
        \node[label] at (4, -1.2) {\(N_n\)};
    \end{tikzpicture}
    \caption{Classification of radiative amplitudes into external (left) and internal (right) emission diagrams. In the case of the external emission diagrams, we denote the non-radiative subdiagram by $\Gamma_n$ or $\tilde{\Gamma}_n$, the off-shell function associated with the incoming or outgoing external fermion leg undergoing the soft emission, respectively.}
    \label{fig:QEDampclass}
\end{figure}

We briefly review the subleading soft theorem in massive QED at tree level \cite{Low:1958sn, Burnett:1967km}. That is, we consider $U(1)$ gauge theory with a single massive Dirac fermion field,
\begin{align}
	\mathcal{L}_\mathrm{QED} = i\psibar D_\mu \gamma^\mu\psi -m\psibar\psi- \frac{1}{4}F^{\mu\nu}F_{\mu\nu}\,,
\end{align}
where $D_\mu = \partial_\mu + ie A_\mu$ is the covariant derivative.
We may split the radiative tree-level amplitude into contributions due to external and internal emissions,
\begin{align}
	\cA_{n+\gamma(q)}  = \sum_{i=1} T^{(i)}_{n}(\{p_j\},q)  + N_{n}(\{p_j\},q),
\end{align}
where the sum above extends over all external fermion legs. Here, $T_n^{(i)}$ and $N_n$ represent diagrams where the soft photon attaches to the $i^\mathrm{th}$ external fermion leg and an internal line, respectively. These contributions are shown diagrammatically in \cref{fig:QEDampclass}.

We orient the momenta of the soft photon to be outgoing and consider scattering involving only incoming/outgoing fermions.
Denoting the set of hard momenta as $\{p_j\}$ and soft photon momenta as $q$, we parameterize the soft limit by rescaling $q \to \tau q$, where $\tau \rightarrow 0$, and hard momenta $p_j$ held fixed. The soft theorem may be expressed as,
\begin{align}
	\lim_{\tau\to 0}\cA_{n+\gamma(\tau q)} = \left(\frac{1}{\tau}\QEDSoftLeading + \QEDSoftSubleading \right)\cA_n\,.
\end{align}
The action of the soft factors $\SoftFactor{i}$ on the lower-point amplitude will depend on the orientation of the external charged states in $\cA_n$.
Due to the presence of the nearly on-shell fermion propagator in $T_{n}^{(i)}$, only external emission diagrams contribute at leading order in the soft scale $\tau$. The remaining subleading soft factor can  then be related to these external emission diagrams using the Ward identity \cite{Low:1958sn, Burnett:1967km}. At tree level, the leading soft factor takes the form \cite{Weinberg:1965nx},
\begin{align}
 \QEDSoftLeading\cA_n &= \sum_\mathrm{out} e\frac{p_i\cdot \epsilon}{p_i\cdot q} \cA_n(\{p_{j}\})-\sum_\mathrm{in} e\frac{p_i\cdot \epsilon}{p_i\cdot q}\cA_n(\{p_j\}),
\end{align}
composed as a sum of contributions from incoming and outgoing fermion external legs, where $\epsilon_\mu$ is the polarization of the soft photon.

While the leading soft behavior may be expressed directly in terms of the lower-point amplitude, the subleading soft factor must be evaluated on the off-shell lower-point function appearing inside $T_n^{(i)}$, before contraction with the external on-shell spinor \cite{Burnett:1967km}. Let us isolate the contribution from a specific ingoing or outgoing external fermion associated with the soft emission. We define off-shell functions $\Gamma_n$ and $\tilde{\Gamma}_n$ by stripping off an external spinor from the lower-point amplitude such that $\cA_{n} = \Gamma_n u(p_i)$ for an ingoing leg and $\cA_n = \overline{u}(p_i)\tilde{\Gamma}_n$ for an outgoing leg. The subleading soft factor may then be expressed as \cite{Burnett:1967km},
\begin{align}
 	\QEDSoftSubleading\cA_n &=  \sum_\mathrm{out} e \overline{u}(p_i)\frac{i\overset{\rightarrow }{\mathcal{J}_{(i)}^{\mu\nu}}q_\mu\epsilon_\nu}{p_i\cdot q}\tilde{\Gamma}_n(\{p_j\}) \nonumber\\
 	 &  \qquad - \sum_\mathrm{in} e \Gamma_n(\{p_j\})\frac{i\overset{\leftarrow }{\mathcal{J}_{(i)}^{\mu\nu}}q_\mu\epsilon_\nu}{p_i\cdot q}u(p_i).
\label{eq:QEDSTF}
\end{align}
The subleading factor is determined by the total angular momentum $\mathcal{J}_{(i)}^{\mu\nu}$ for the $i^\mathrm{th}$ external leg,
\begin{align}
	\mathcal{J}_{(i)}^{\mu\nu} =
	\mathcal{J}_{(i),\mathrm{orb}}^{\mu\nu} + i[\gamma^\mu, \gamma^\nu]/4\,,
\end{align}
where
\begin{align}
	\mathcal{J}_{(i),\mathrm{orb}}^{\mu\nu}
	= i\left(p_{i}^{\mu}\frac{\partial}{\partial p_{i}^\nu} - p_{i}^{\nu}\frac{\partial}{\partial p_{i}^\mu}\right).
    \label{eq:ang-mom-qed}
\end{align}
Here $\mathcal{J}_{(i),\mathrm{orb}}^{\mu\nu}$ is the orbital generator, while the second term is the spin generator.
The derivatives with respect to external hard momenta $p_i$ in \cref{eq:QEDSTF} should be understood to act only on $\Gamma_n$ or $\tilde{\Gamma}_n$.
For compactness, we will adopt the notation prevalent in the literature to write \cref{eq:QEDSTF} as
\begin{align}
	\QEDSoftSubleading\cA_n = \sum_i \eta_i e \frac{i{\mathcal{J}_{(i)}^{\mu\nu}}q_\mu\epsilon_\nu}{p_i\cdot q}\cA_n(\{p_j\})\,,
\end{align}
where $\eta_i = \mp 1$ for incoming/outgoing external fermion legs, respectively.

Let us now pass to the context of a general EFT: a massive fermion coupled to a $U(1)$ gauge field with the same global symmetries as QED. While the leading soft factor is universal, the tree level subleading soft factor is sensitive to higher-dimensional operators in the EFT \cite{Elvang:2016qvq}. More precisely, there exists only a single dimension-five operator that modifies the subleading soft photon theorem at tree level. This is the Pauli operator,
\begin{align}
\mathcal{O}_P = i\psibar \sigma^{\mu\nu}F_{\mu\nu}\psi,
\label{eq:pauli-operator}
\end{align}
where $\sigma^{\mu\nu} = i [\gamma^\mu, \gamma^\nu]/2$.
We denote QED supplemented by the Pauli operator as
\begin{align}
\mathcal{L}_{\mathrm{QED+P}} = \mathcal{L}_\mathrm{QED} + \frac{\Cpauli}{\Lambda}\mathcal{O}_P,
\label{eq:QEDdeform}
\end{align}
where $\Lambda$ is a heavy scale suppressing the dimensionless Wilson coefficient $\Cpauli$ of the Pauli operator. 
Because the soft photon enters through $F_{\mu\nu}$, the Pauli operator vertex carries an explicit power of the soft momentum and cannot modify the leading soft factor. Its presence in the EFT induces a new local transverse interaction vertex which modifies the subleading soft factor to
\begin{align}
	\QEDSoftSubleadingPauli\cA_n = &\sum_i \eta_i  \left[e\mathcal{J}_{(i)}^{\mu\nu}-2\frac{\Cpauli m}{\Lambda} \sigma^{\mu\nu}\right]\frac{iq_\mu\epsilon_\nu}{p_i\cdot q}\cA_n\,.
\label{eq:QEDEFTtree}
\end{align}
The Pauli operator modifies the asymptotic magnetic moment of the external fermions \cite{Aebischer:2021uvt}. The new term is gauge-invariant, and thus is not related to the leading soft factor by gauge-invariance. Note that we specifically consider fermionic QED, and its EFT generalization. In the case of an EFT for scalar QED, no such gauge invariant higher-dimensional operator, consistent with Lorentz symmetry, exists at sufficiently low mass dimension to modify the subleading soft photon theorem.

At one loop, the leading soft factor is known to be universal in QED, while the subleading soft factor receives $\log \tau$ corrections \cite{Engel:2021ccn, Engel:2023ifn, Sahoo:2018lxl, Krishna:2023fxg, AtulBhatkar:2018kfi, Laddha:2018myi}. In a massive theory such as QED, corrections of this type that are non-analytic in external kinematics may only be generated by soft or collinear integration regions. 
By contrast, the hard integration region can only generate corrections to terms analytic in the external kinematics. In minimal QED, this region contributes only finite terms to the subleading soft factor. As we will find, this conclusion is modified in an EFT that includes higher-dimensional operators. The Pauli operator supplies a local transverse interaction vertex that, unlike the QED interaction vertex, receives UV divergent corrections from the hard integration region that manifest in the subleading soft factor.

We therefore next consider one loop corrections to the soft theorem in the EFT \cite{Engel:2021ccn, Sahoo:2018lxl, Krishna:2023fxg, AtulBhatkar:2018kfi, Engel:2023ifn, Bonocore:2014wua}. We distinguish the hard integration region to be where the loop momentum $\ell$ is of the same order as the fermion masses and hard external invariants, and the soft integration region to be where $\ell \sim \tau q$ scales with the emitted photon momentum. Note that since the external fermion masses are held fixed in the soft limit, there is no separate collinear region in this power counting. This differs from the massless SCET setting, where on-shell collinear configurations modify the naive Low-Burnett-Kroll theorem at subleading order \cite{Larkoski:2014bxa}.

At one loop, the Ward identity argument leading to  \cref{eq:QEDSTF} is insufficient to distinguish external and internal emission contributions at leading and subleading order. The soft integration region of the internal-emission diagrams can -- in naive power counting -- contribute at order $\mathcal{O}(\tau^{-1})$ and $\mathcal{O}(\tau^{0})$ in the soft theorem. Hence, one cannot deduce internal-emission contributions from soft power-counting and gauge-invariance alone. In \cite{Engel:2021ccn} a classification of the one loop Feynman diagrams contributing to the soft region has been performed, leading to the $\mathcal{O}(\log \tau)$ contribution to the subleading soft theorem. 

The contribution from the soft integration region is known to be unmodified by the presence of higher-dimensional operators to subleading order \cite{Krishna:2023fxg, AtulBhatkar:2018kfi, Sahoo:2018lxl}. We therefore turn our focus to contributions from the hard integration region. Since the hard integration region does not extend down to the soft scale, it cannot generate any structures that are non-analytic in $\tau$. 
Thus, the usual Ward identity argument utilized in obtaining \cref{eq:QEDSTF} applies here.

The form of the leading soft factor is unmodified\footnote{The charge $e$ appearing in \cref{eq:QEDEFTLSF} and subsequent equations corresponds to the physical electromagnetic charge \cite{Bogolyubov:1956gh}; see \cref{sec:appendix-qed-examples}.}
\begin{align}
	\QEDSoftLeading\cA_n = \sum_i  \eta_i e\frac{p_i\cdot \epsilon}{p_i\cdot q} \cA_n \,,
\label{eq:QEDEFTLSF}
\end{align}
which is a consequence of a Ward-Takahashi identity (WTI) for the $U(1)$ global symmetry \cite{Weinberg:1964ew, Bern:2014vva, Laddha:2017vfh, DelDuca:1990gz}.

As shown in \cref{sec:appendix-qed-examples}, the presence of the higher dimensional operator induces UV-divergent loop corrections to the subleading soft factor.\footnote{We omit the IR divergences in the following discussion as their analysis involves the soft integration region contributions considered in \cite{Engel:2021ccn,Krishna:2023fxg}.} These contributions to the three-point vertex are both unconstrained by the WTI and survive in the soft limit. Including the contributions from the hard integration region, the subleading soft factor can be written as
\begin{align}
	\QEDSoftSubleading_{\text{finite}}\cA_n = &\sum_i \eta_i   \left[e\mathcal{J}_{(i),\mathrm{orb}}^{\mu\nu}+F_M  \sigma^{\mu\nu}\right]\frac{iq_\mu\epsilon_\nu}{p_i\cdot q}\cA_n\,,
\label{eq:QEDSLSF}
\end{align}
where $\mathcal{J}_{(i),\mathrm{orb}}^{\mu\nu}$ is the orbital angular momentum operator defined through \cref{eq:ang-mom-qed} and $F_M$ parametrizes the spin-dependent contributions.

Working within the modified minimal subtraction ($\overline{\rm MS}$) scheme for the theory defined by $\mathcal{L}_{\mathrm{QED+P}}$ in \cref{eq:QEDdeform}, $F_M$ has the form
\begin{align}
	F_M &= \frac{e}{2}- 2\frac{\Cpauli m_p}{\Lambda} + \eta(e,\Cpauli)\frac{m_p}{\Lambda}\log \frac{\mu^2}{m_p^2} +\xi\,,
\label{eq:QEDEFTFis}
\end{align}
where
\begin{equation}
    \eta(e,C) = \frac{C}{24\pi^2}\left(60 \frac{\Cpauli^2m_p^2}{\Lambda^2} -66 e \frac{\Cpauli m_p}{\Lambda}+ 17e^2\right).
\end{equation}
Here, $C(\mu)$ is the renormalized coupling, $\mu$ is an arbitrary renormalization scale, $m_p$ is the physical pole mass of the fermions,  and  $\xi$ is a UV-finite scheme-dependent factor.

Since the amplitudes $\cA$ must be independent of the renormalization scale $\mu$ for any number of external states $n$,
\begin{align}
	\mu \frac{d}{d\mu}\cA_n = 0\,,
\end{align}
from the soft theorem we can deduce that the soft factors relating them must be independent of $\mu$ as well. They therefore satisfy,
\begin{align}
	\mu \frac{d}{d\mu}\SoftFactor{i} = 0\,.
\label{eq:SFRG}
\end{align}
For the subleading soft factor in \cref{eq:QEDSLSF}, the separation into unique tensor structures indicates the scalar coefficient of each must be independently RG invariant. For the orbital angular momentum operator $\mathcal{J}_{(i),\mathrm{orb}}^{\mu\nu}$, this is satisfied by the appearance of the physical electromagnetic charge $e$. For $F_M$, \cref{eq:SFRG} implies the running of the Pauli operator according to,
\begin{equation}
    \mu\frac{d}{d\mu}C(\mu) = \eta(e(\mu),C(\mu))\,,
\end{equation}
consistent with the results of \cite{Jenkins:2017dyc}.
Solving this equation allows one to evolve the couplings in the soft factors down to an intermediate scale $\mu\sim m_p$, which renders the logs in \cref{eq:QEDEFTFis} small. Here the fermion mass provides a natural matching scale for an EFT describing infrared dynamics, which may be used to evaluate the remaining contributions from the soft integration region \cite{Engel:2023ifn, Engel:2023rxp}.
The soft theorem is then most naturally expressed in terms of parameters evaluated at this matching scale. Including the contributions from the soft integration regions, the soft theorem takes the form,
\begin{align}
\lim_{\tau\to 0}\cA_{n+\gamma(\tau q)}
    = \Big(\frac{1}{\tau}\QEDSoftLeading + \log \tau\, \QEDSoftSubleadingLog
    + \QEDSoftSubleadingFinite \Big)\cA_n\,, \nonumber
\end{align}
where the coefficient $C$ of the Pauli operator is evaluated at the mass scale $\mu = m_p$. The exact form of $\QEDSoftSubleadingLog$ can be found in \cite{Engel:2021ccn}.
Universality of the subleading soft photon theorem in $\mathcal{L}_{\mathrm{QED+P}}$ is restored at one loop by virtue of this $\log \tau$ term  \cite{Krishna:2023fxg, Sahoo:2018lxl, AtulBhatkar:2018kfi}. The non-universal dependence on the higher-dimensional operator is entirely relegated to $\QEDSoftSubleadingFinite$, which is subdominant in the soft limit.

A natural question is how these results apply to theories involving massless fermions. The absence of a fermion mass leaves collinear singularities unsuppressed \cite{Ma:2023gir, DelDuca:1990gz,Vernazza:2023hrf}, introduces loop corrections to the leading soft factor \cite{Ma:2023gir}, complicates factorization \cite{Laenen:2020nrt, Bern:2014oka,Vernazza:2023hrf}, and fundamentally alters the soft loop contributions \cite{Yennie:1961ad, Becher:2009cu}. The EFT generalization of the soft theorem will also be modified. In contrast to the massive theory, where the addition of a higher-dimensional operator modifies the subleading soft factor, deformations of massless QED are further constrained by a non-invertible symmetry associated with axial rotations of the fermions by rational angles \cite{Choi:2022jqy,Cordova:2022ieu}. Although the $U(1)_A$ symmetry of the action is broken by the ABJ anomaly, this resulting non-invertible symmetry acts on local operators, thereby forbidding chirality-violating operators in the EFT, such as the Pauli operator. Assuming the infrared theory respects this symmetry, there are no candidate UV deformations with appropriate dimensions to modify the subleading soft factor. Deforming massless QED by a Pauli operator explicitly breaks the global axial symmetry and a bare fermion mass term is no longer forbidden; rather, naturalness arguments necessitate its inclusion. Consequently, there is no EFT generalization for the subleading soft factor in a theory involving massless fermions.

\section{Soft pions}\label{sec:chiraleft}
The second example of a running soft theorem we consider is the double-soft theorem for pion scattering amplitudes in the massless Chiral EFT.
Soft-pion theorems have been studied extensively \cite{Adler:1964um,Weinberg:1966gjf,Arkani-Hamed:2008owk,Kampf:2013vha,Cachazo:2015ksa,Du:2015esa,Rodina:2021isd}, including one-loop analyses of single-soft behavior at the integrand level \cite{Bartsch:2022pyi,Bartsch:2024ofb}.
The leading double-soft theorem, which encodes the NLSM current algebra, was originally derived by Weinberg \cite{Weinberg:1966gjf}.
Subleading extensions were later obtained in \cite{Kampf:2013vha,Cachazo:2015ksa,Du:2015esa,Rodina:2021isd}.
More recently, \cite{Rodina:2021isd} derived the tree-level double-soft theorem in the Chiral EFT and showed that the only corrections arise from four-derivative interactions.
Here we extend this analysis to one loop, where loop effects enter at the same order in chiral power counting as four-derivative interactions.
We work with flavor-ordered amplitudes in this section and give the full flavor-dressed theorem in \cref{sec:appendix-full-double-soft-pions}.
The result shows that universal logarithms dominate the subleading soft behavior, while the remaining finite terms are controlled by running Wilson coefficients.

We start by reviewing the tree-level subleading double-soft theorem in the Chiral EFT with massless $SU(N)$ pions. The chiral Lagrangian is organized as a derivative-expansion,
\begin{align}
    \mathcal{L}_{\chi \rm EFT} = \mathcal{L}^{(2)} + \mathcal{L}^{(4)}+\cdots\,. \label{eq:chiral-lagrangian}
\end{align}
The two-derivative term $\mathcal{L}^{(2)}$ is the NLSM Lagrangian, which can be parametrized by
\begin{align}
    \mathcal{L}^{(2)} &= \frac{f^2}{4}\langle \partial_\mu U \partial^\mu U^{-1} \rangle\,, \label{eq:lagrangian-nlsm}
\end{align}
where $f$ is the pion decay constant and
\begin{align}
    U(x) = \exp\left(\frac{i \sqrt{2}}{f}\pi^a(x) t^a\right).
\end{align}
Here $\pi(x)^a$ is the pion field, $t^a$ are the generators of $SU(N)$ normalized by $\langle t^a t^b \rangle = \delta^{ab}$ and we introduced the notation $\langle \dotsm \rangle \equiv \mathrm{Tr}(\dotsm)$ for the trace over flavor indices.

At four-derivative order $\mathcal{L}^{(4)}$, for $SU(N)$ there are four independent operators \cite{Gasser:1983yg} which we choose as
\begin{align}
    \mathcal{L}^{(4)} =
     \,&L_1 \langle \partial_\mu U^\dagger \partial^\mu U \rangle
          \langle \partial_\nu U^\dagger \partial^\nu U \rangle \nonumber\\
    + &L_2 \langle \partial_\mu U^\dagger \partial_\nu U \rangle
          \langle \partial^\mu U^\dagger \partial^\nu U \rangle \nonumber\\
    + &L_3 \langle \partial_\mu U^\dagger \partial^\mu U \partial_\nu U^\dagger \partial^\nu U \rangle \nonumber\\
    + &L_4 \langle \partial_\mu U^\dagger \partial_\nu U \partial^\mu U^\dagger \partial^\nu U \rangle\,,
    \label{eq:chiral-lagrangian-in-text} 
\end{align}
controlled by Wilson coefficients $L_i$. By derivative-counting arguments, operators with derivative number higher than four will not modify the subleading double-soft factor \cite{Rodina:2021isd}, so we suppress them here.

Recall that the planar single-trace sector of pion amplitudes admits a decomposition into \textit{flavor-ordered} partial amplitudes.
For a scattering amplitude with $n$ pions we write
\begin{align}
    \cA_n
    &=
    \sum_{\sigma \in S_n/Z_n}
    \langle t^{a_{\sigma(1)}}\dotsm t^{a_{\sigma(n)}}\rangle\,
    A_n(p_{\sigma(1)},\ldots,p_{\sigma(n)})\,. \label{eq:flavor-ordering-at-tree-level}
\end{align}
Then the double-soft theorem \cite{Weinberg:1966gjf,Kampf:2013vha,Cachazo:2015ksa} can be stated at the level of flavor-ordered partial amplitudes. In what follows, we focus only on the case where the two soft particles are adjacent to one another in the ordering. The soft behavior of all other orderings, as well as multiple-trace amplitudes, can straightforwardly be derived from the full flavor-dressed double-soft theorem discussed in \cref{sec:appendix-full-double-soft-pions}.

We parameterize the double-soft limit by scaling the soft momenta as $q_a\to\tau q_a$ and $q_b\to\tau q_b$ with $\tau\to0$, while keeping all hard external momenta fixed. With $p_L$ and $p_R$ denoting the hard momenta adjacent to the soft pair, as in \cref{fig:double_soft_limit}, the theorem takes the schematic form
\begin{align}
    &\lim_{\tau\rightarrow 0}
    A_{n+2}(\ldots,p_L,\tau q_a,\tau q_b,p_R,\ldots) \label{eq:double-soft-theorem-schematic-nlsm}\\
    &\qquad =
    \left(\NLSMSoftLeading + \tau \NLSMSoftSubleading \right)
    A_n(\ldots,p_L,p_R,\ldots)\,. \nn
\end{align}
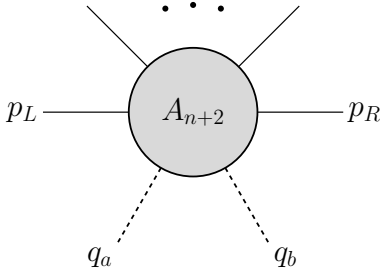
\begin{figure}[t]
    \centering
    \resizebox{0.6\columnwidth}{!}{
        \begin{tikzpicture}[thick]

            \draw[fill=gray!30, draw=black, very thick] (0,0) circle (1.5cm);
            \node at (0,0) {\huge $A_{n+2}$};


            \draw (0:1.5) -- (0:3.5) node[right, font=\huge] {$p_R$};

            \draw (180:1.5) -- (180:3.5) node[left, font=\huge] {$p_L$};

            \draw (45:1.5) -- (45:3.5);
            \draw (135:1.5) -- (135:3.5);

            \fill (75:2.5) circle (2pt);
            \fill (90:2.5) circle (2pt);
            \fill (105:2.5) circle (2pt);


            \draw[dashed, black, very thick] (240:1.5) -- (240:3.5) node[below left, text=black, font=\huge] {$q_a$};

            \draw[dashed, black, very thick] (300:1.5) -- (300:3.5) node[below right, text=black, font=\huge] {$q_b$};

        \end{tikzpicture}
    }
    \caption{The double-soft theorem of \cref{eq:double-soft-theorem-schematic-nlsm} corresponds to the simultaneous soft limit of the momenta $q_a$ and $q_b$ in the ordered partial amplitude $A_{n+2}(\ldots,p_L,\tau q_a,\tau q_b,p_R,\ldots)$.}
    \label{fig:double_soft_limit}
\end{figure}
The leading double-soft factor $\NLSMSoftLeading$ is \cite{Weinberg:1966gjf}
\begin{align}
    \NLSMSoftLeading = -\frac{1}{4f^2}\left(
        \frac{p_L\cdot (q_a-q_b)}{p_L\cdot q_{ab}}
        + (p_L,q_a) \leftrightarrow (p_R,q_b)
    \right), \label{eq:leading-double-soft-factor-nlsm}
\end{align}
where $q_{ab}\equiv (q_a + q_b)$. The leading soft factor is universal in the sense that it is not modified by loop corrections and higher-derivative operators in the Chiral EFT, which follows from symmetry current algebra \cite{Weinberg:1966gjf, Kampf:2013vha}.
However, such corrections do enter at the next order in the soft expansion.

The subleading soft factor $\NLSMSoftSubleading$ in  \cref{eq:double-soft-theorem-schematic-nlsm} has the form \cite{Cachazo:2015ksa,Rodina:2021isd,Berean-Dutcher:2025ohp}
\begin{align}
    \NLSMSoftSubleading &=
      -\frac{1}{2f^2}\bigg(
                \frac{(q_a\cdot q_b)\,p_L\cdot q_b}{\left(p_L\cdot q_{ab}\right)^2}
              + (p_L,q_a) \leftrightarrow (p_R,q_b)
            \bigg) \nn\\
    &\quad - \frac{1}{2f^2}
        \bigg(
             \frac{q_a^\mu q_b^\nu}{p_L\cdot q_{ab}}\mathcal{J}^{(L)}_{\mu\nu}
            +\frac{q_b^\mu q_a^\nu}{p_R\cdot q_{ab}} \mathcal{J}^{(R)}_{\mu\nu}
        \bigg) \nn\\
    &\quad +\frac{1}{2f^4}
        \bigg(
              \frac{(p_L\cdot q_a)(p_L\cdot q_b)}{p_L \cdot q_{ab}}\mathcal{F}(p_L,q_a;\tau)\nn\\
    &          \hspace{2.8cm}+ (p_L,q_a) \leftrightarrow (p_R,q_b)\bigg)\,, \label{eq:sub-leading-double-soft-factor-nlsm}
\end{align}
where $\mathcal{J}^{(i)}_{\mu\nu}$ denotes the angular momentum operator acting on the $i^\mathrm{th}$ external pion. For scalar external states, it is the momentum-derivative operator appearing in \cref{eq:ang-mom-qed}.
The function $\mathcal{F}(p_L,q_a;\tau)$ parameterizes the contribution from higher-derivative operators \cite{Rodina:2021isd} and loops; see \cref{eq:f-correction-to-soft-factor} for the full expression.

We obtained the double-soft theorem at subleading order by following the current algebra derivation in \cite{Berean-Dutcher:2025ohp}. In this approach, $\NLSMSoftSubleading$ is derived from the form factors with current insertions in the soft limit, which obey the symmetry current algebra.\footnote{Alternatively, one could attempt to derive the subleading soft factor using diagrammatic arguments \cite{Du:2015esa,Low:2015ogb,Cohen:2025dex}. The lack of a three-point vertex simplifies the analysis in method of regions as compared to QED.} The coefficients in the first two lines in \cref{eq:sub-leading-double-soft-factor-nlsm} are fixed entirely by imposing the constraints from chiral symmetry \cite{Berean-Dutcher:2025ohp}. In contrast, to capture the one loop corrections and non-universal dependence on Wilson coefficients, we compute the relevant soft form factors with current insertions from a particular parametrization of the Chiral EFT Lagrangian given by \cref{eq:chiral-lagrangian,eq:lagrangian-nlsm,eq:chiral-lagrangian-in-text}. Finally, note that employing power-counting arguments as in \cite{Rodina:2021isd} for loops in the Chiral EFT, we can deduce that contributions from two loops and higher will not modify the subleading soft factor, thus \cref{eq:sub-leading-double-soft-factor-nlsm} is one loop exact.

We package the correction to the soft factor in a scheme-independent way \cite{Guerrieri:2020bto} as
\begin{align}
    \mathcal{F}(p_L,q_a;\tau)
    &=\alpha + \frac{N}{48\pi^2}\log\frac{-f^2}{2\,p_L\cdot q_a\,\tau}\,, \label{eq:f-correction-to-soft-factor}
\end{align}
where $\alpha$ is a physical parameter defined at the scale $f$. 
As for an explicit expression for $\alpha$ in $\overline{\rm MS}$ scheme, using the operator basis defined in \cref{eq:chiral-lagrangian-in-text} yields
\begin{align}
    \alpha = 8\left(L_3^r(\mu)+2L_4^r(\mu)\right)
    + \frac{N}{48\pi^2}\left(\frac{13}{6} + \log\frac{\mu^2}{f^2}\right), \label{eq:appendix-alpha-li-relation}
\end{align}
where $L_i^r(\mu)$ are the renormalized Wilson coefficients.

Note the presence of a logarithm that depends on the soft parameter $\tau$ in \cref{eq:f-correction-to-soft-factor}. To discuss the implications for universality, we group the terms in the double-soft theorem \cref{eq:double-soft-theorem-schematic-nlsm} as
\begin{align}
    \lim_{\tau\rightarrow 0} A_{n+2}
    = \left(\NLSMSoftLeading + \tau\log\tau\,\NLSMSoftSubleadingLog + \tau \NLSMSoftSubleadingFinite\right)A_{n}\,. \label{eq:double-soft-univ-non-univ}
\end{align}
Here $\NLSMSoftSubleadingLog$ isolates the coefficients of the soft logarithms, which are independent of the higher-derivative operators. The non-universal correction $\alpha$ enters via a contribution to $\NLSMSoftSubleadingFinite$ and must be fixed by matching.
Thus, the dominant subleading double-soft pion behavior is universal because of the $\mathcal{O}(\tau \log \tau)$ loop contribution, while the finite local terms remain sensitive to the $\mathcal{O}(p^4)$ Wilson coefficients of the Chiral EFT. 

As explained in the introduction, the subleading soft factor is RG invariant as a whole. This implies
\begin{align}
    \mu \frac{d}{d\mu}\NLSMSoftSubleadingLog = - \mu \frac{d}{d\mu}\NLSMSoftSubleadingFinite\,.
\end{align}
This relation is realized by the running of the $\mathcal{O}(p^4)$ Wilson coefficients \cite{Gasser:1983yg},
\begin{align}
    \mu\frac{d}{d\mu}\big(L_3^r(\mu)+2L_4^r(\mu)\big) = -\frac{N}{192\pi^2}\,,\label{eq:pion-wilson-coefficient-running}
\end{align}
within $\mathcal{F}(p_L,q_a;\tau)$, which we restate here at an arbitrary scale in the $\overline{\rm MS}$ scheme
\begin{align}
    \mathcal{F}(p_L,q_a;\tau)
    &=8\left(L_3^r(\mu)+2L_4^r(\mu)\right) \notag\\
    &\quad
        + \frac{N}{48\pi^2}\left(\frac{13}{6} + \log\frac{-\mu^2}{2\,p_L\cdot q_a\,\tau}\right). \label{eq:conventional-f-function}
\end{align}
We note that in \cref{eq:conventional-f-function}, the logarithms are minimized by choosing the RG scale locally as $\mu^2 \sim \tau |2\,p\cdot q|$, where $p$ and $q$ denote the adjacent hard and soft momenta in the term under consideration.

\section{Conclusion}

In this work, using soft photons and double-soft pions as examples, we showed that subleading soft theorems in EFT are subject to RG evolution.
The leading soft behavior remains fixed by infrared symmetries, while at one loop the subleading terms separate into universal logarithms and finite analytic contributions controlled by running Wilson coefficients.

More broadly, our results suggest that EFT provides the natural organizing principle for subleading corrections to the soft behavior of scattering amplitudes.
In theories where the leading soft limit is fixed by an infrared symmetry, subleading soft corrections have a constrained structure: they separate into universal non-analytic terms determined by EFT loop integrals and finite analytic terms controlled by local Wilson coefficients.
This distinction points toward a general classification of soft limits according to which parts are symmetry protected, which parts are fixed by universal non-analytic structure, and which parts depend on non-universal short-distance EFT data.

It would be interesting to connect this perspective with recent approaches to soft geometric theorems at one loop order \cite{Cohen:2025dex,Cohen:2026xnr} for scalar EFTs without assuming a global symmetry. Although the soft structures in a general case are considerably more involved, one would still expect that various terms can be repackaged using RG running, for example into (derivatives of) self-energy or grouped according to operator-mixing under RG. In particular, the running soft factors studied here may admit an interpretation in terms of the renormalization of field-space geometry \cite{Alonso:2016oah,Alonso:2017tdy,Helset:2022pde,Assi:2023zid,Aigner:2025xyt,Assi:2025fsm}, suggesting a broader framework for understanding soft theorems beyond the specific examples considered in this work.

\begin{acknowledgments}
 We would like to thank Julio Parra-Martinez, Jasper Roosmale Nepveu, Gordon W. Semenoff, and Chia-Hsien Shen for useful discussions. We also thank Callum Jones, Julio Parra-Martinez, Chia-Hsien Shen and Jaroslav Trnka for comments on the manuscript. JBD is supported by a Four-Year Fellowship (4YF) from the University of British Columbia.
MD is funded by the European Union under ERC Starting Grant AmpEFT 101165689. Views and opinions expressed are however those of the authors only and do not necessarily reflect those of the European Union or the European Research Council. Neither the European Union nor the granting authority can be held responsible for them.
RS acknowledges the support of the Natural Sciences and Engineering Research Council of Canada (NSERC). 
\end{acknowledgments}

\appendix
\section{Details of the soft factor in QED}\label[appendix]{sec:appendix-qed-examples}

Here, we construct contributions from the hard integration region to the leading and subleading soft factors. As explained in the main text, this contribution can be obtained purely from the external emission diagrams and application of the LBK theorem. In the local field basis presented in $\mathcal{L}_\mathrm{QED+P}$, the relevant diagrams decompose into the form,
\begin{equation}
    T_{n+\gamma} = \sum_\mathrm{in}\Gamma_n(\{p_i\},q)iD(p_i-q)iV^\mu(p_i,q)  N_\gamma \epsilon_{\mu}u(p_i) + \ldots 
\label{eq:QEDextdia}
\end{equation}
where $\Gamma_n$ encodes the lower-point hard process (see \cref{fig:QEDampclass}), $V^\mu(p,q)$ is the three-point interaction vertex, $iD(p)$ is the fermion propagator, and $\epsilon_\mu$ denotes the external soft photon polarization. Ellipses denote contributions from outgoing external fermion states which follow similarly as to the ingoing case. We will neglect these below for brevity. The factor $N_\gamma$ captures the overlap with the local photon field $A_\mu(q)$ and the asymptotic one-photon state,
\begin{equation}
    \langle\Omega| A^\mu(q)|\gamma_h(q) \rangle = N_\gamma\epsilon_h^\mu(q)\,,
\end{equation}
where $h$ is the helicity label, which we will suppress in what follows.

The factor $N_\gamma$ is present due to the construction of physical amplitudes from off-shell correlation functions via LSZ reduction. This process involves projecting asymptotic states out of the local fields present in the correlation functions, introducing overlap factors for each external leg. Anticipating the factorization of $T_{n+\gamma}$ into a lower point amplitude, we have made the overlap factor associated with the soft photon explicit, while the analogous factors for the hard legs are included in $\Gamma_n$.

We may then proceed by characterizing the quantities that appear in \cref{eq:QEDextdia}. The fermion propagator $iD(p)$ may be expressed as,
\begin{align}
	iD(p) = \frac{i}{\slashed{p} - m + \Sigma(p)+i\epsilon}\,,
\end{align}
where the self-energy $\Sigma = \Sigma_1(p)\slashed{p} -\Sigma_2(p)$ captures loop corrections. The overlap factor $N_\gamma$ is determined by the residue of the photon propagator at the on-shell pole,
\begin{equation}
    \lim_{q^2\rightarrow 0 }\langle TA_{\mu}(q) A_{\nu}(-q) \rangle = \frac{-iN_\gamma^2}{q^2 + i\epsilon}\Pi^{\mu\nu}+\ldots,
\end{equation}
where $\Pi^{\mu\nu}$ encodes the $\mathcal O(q^0)$ Lorentz structure, and ellipses denote contributions from multi-particle states which begin at the mass threshold $q^2 \sim m_p^2$. The three-point interaction vertex may be decomposed into longitudinal and transverse parts,
\begin{align}
	V^\mu(p,q) = V^\mu_\mathrm{L}(p,q) + V^\mu_\mathrm{T}(p,q)\,,
\end{align}
where $q\cdot V_\mathrm{T} = 0$.
The Ward-Takahashi identity relates $V_\mathrm{L}$ above to the fermion propagator,
\begin{align}
	q\cdot V_\mathrm{L} = e\left[D^{-1}(p-q) - D^{-1}(p) \right]\,.
\label{eq:QEDWTI}
\end{align}
Ignoring effects from the soft integration region, we may use the above identity to express $V_\mathrm{L}$ as a series expansion,
\begin{align}
	V_\mathrm{L}^\mu u(p)= e\sum_{n=0}^\infty \frac{(-1)^{n+1}}{n!}q_{\nu_1}... \ q_{\nu_{n}}\frac{\partial^{n+1}D^{-1}(p)}{\partial p_\mu \partial p_{\nu_1}... \partial p _{\nu_n}}u(p)\,.
\end{align}
By contrast, the transverse vertex $V_\mathrm{T}$ is unconstrained.
We parameterize its general form as
\begin{align}
V_\mathrm{T}^\mu(p,q)u(p)\epsilon_\mu = \left[\sum_j w_j(p\cdot q)K_j^\mu\right] u(p)\epsilon_\mu\,,
\end{align}
where $w_j$ are scalar functions of the momenta and $K_j^\mu \in \{\gamma^\mu, p^\mu, \gamma^\mu \slashed{q}, p^\mu\slashed{q} \}$ are possible independent spin structures.
Imposing the transverse condition $q \cdot V_\mathrm{T}=0$, we find,
\begin{align}
	V_\mathrm{T}^\mu u(p)\epsilon_\mu= \left[w_1\left(p^\mu \slashed{q}-p\cdot q \gamma^\mu\right) + w_2 \gamma^\mu \slashed{q}\right]u(p)\epsilon_\mu\,.
\end{align}
The transverse condition implies that $V_\mathrm{T}$ begins at subleading order.

Using this parameterization of the three-point vertex, we derive an explicit form for the contributions from the hard integration region to the soft theorem,
\begin{align}
	\lim_{\tau\to 0}\cA_{n+\gamma(\tau q)} = \left(\frac{1}{\tau}\QEDSoftLeading + \QEDSoftSubleading + \mathcal{O}(\tau)\right)\cA_n\,.
\end{align}
The leading soft factor appears as,
\begin{equation}
    S^{(-1)}\mathcal{A}_n = \sum_i \eta_i  \left(eN_\gamma\right)\frac{p_i \cdot \epsilon}{p_i \cdot q} \mathcal{A}_n\,,
\label{eq:QEDLSFapp}
\end{equation}
where the combination $eN_\gamma$ is a scheme-independent, physical coupling \cite{Bogolyubov:1956gh}. In terms of the photon vacuum polarization $\Pi(q^2)$ it takes the familiar form,
\begin{equation}
    eN_\gamma = \lim_{q^2\rightarrow 0}\frac{e}{\sqrt{1+\Pi(q^2)}}\,.
\end{equation}
The hard-region contribution to the subleading soft factor \cref{eq:QEDSLSF} is parametrized by $F_M$, evaluated here in terms of the fermion pole mass $m_p$,
\begin{align}
			F_M &=N_\gamma \frac{-2m_p w_2(0)+e \Delta_1(m_p)}{2(\Delta_1(m_p) + 2m_p \Delta_2(m_p))}\,,
\end{align}
where
\begin{align}
	\Delta_1(m_p) = 1 + \Sigma_1(m_p), \quad \Delta_2(m_p) = m_p\Sigma_1^\prime - \Sigma_2^\prime\,,
\end{align}
and $\Sigma_n^\prime \equiv \partial_{p^2} \Sigma_n$ is evaluated on-shell $p^2=m_p^2$.
Notice the scalar coefficient $w_1$ does not appear in the subleading soft factor.
Since the role of the hard integration region is to contribute to local Wilson coefficients in the effective action, this restates that only a deformation of the form in \cref{eq:QEDdeform} can modify the subleading soft theorem \cite{Elvang:2016qvq}.
Keeping only one loop contributions, we obtain
\begin{align}
			F_M &= N_\gamma\left[\frac{e}{2} - m_p w_2(0)\left(2-\Delta_1 - 2 m_p\Delta_2\right)-e m_p \Delta_2\right]. 
\end{align}
Including the necessary counterterms, explicit calculation in the EFT described by $\mathcal{L}_{\mathrm{QED+P}}$ yields \cref{eq:QEDEFTFis} in a $\overline{\text{MS}}$ scheme.

It is illuminating to compare the form of the soft factors expressed in different schemes. In an on-shell scheme, the photon overlap factor $N_\gamma$ is set to 1 and the charge $e$ appearing in \cref{eq:QEDLSFapp} corresponds directly to the physical coupling. Conversely, in a different scheme such as $\overline{\text{MS}}$, the overlap factor $N_\gamma$ retains explicit scale dependence. In the leading soft factor one finds,
\begin{equation}
    \sum_i \eta_i \left(\frac{e(\mu)}{\sqrt{1+\overline{\Pi}(q^2=0, e(\mu), C(\mu))}} \right)\frac{p_i\cdot \epsilon}{p_i\cdot q} \mathcal{A}_n\,,
\end{equation}
where $e(\mu)$, $C(\mu)$, and $\overline{\Pi}(\mu)$ correspond to the renormalized couplings and photon vacuum polarization in the $\overline{\text{MS}}$ scheme. The soft factors themselves are RG scale-invariant, which in $\overline{\text{MS}}$ is realized by the running of the renormalized charge
\begin{equation}
    \mu \frac{d}{d\mu}e(\mu) = e(\mu)\frac{\gamma_A}{2}\,,
\end{equation}
where
\begin{align}
    \gamma_A=\frac{1}{1+\overline{\Pi}}{\,}\mu \,\frac{d}{d\mu}\overline{\Pi}
\end{align}
is the anomalous dimension of the photon field $A_\mu$ in this scheme. The same relation can be shown to follow from the $U(1)$ WTI as presented in \cref{eq:QEDWTI}. At one loop order in $\mathcal{L}_\mathrm{QED+P}$, a direct calculation yields,
\begin{equation}
    \gamma_A = \frac{1}{6\pi^2}\left(e^2 -12 e\frac{\Cpauli m_p}{\Lambda} +12 \frac{\Cpauli ^2m_p^2}{\Lambda^2}\right).
\end{equation}

Finally, it is instructive to contrast this running with the case of the double-soft pion theorem. No such single-particle matrix element contributions appear in the running of the $L_i$ Wilson coefficients in \cref{eq:pion-wilson-coefficient-running}. This can be understood as a consequence of symmetry and power counting in the Chiral EFT. In the derivation of the double-soft pion theorem from current algebra, the axial current $J_\mu^a$, associated to the non-linearly realized $SU(N)_A$ transformations \cite{Kampf:2013vha}, is the interpolating operator with a one-pion matrix element. Symmetry fixes the structure of this matrix element,
\begin{align}
    \langle \Omega | J_\mu^a(0) | \pi^b(q) \rangle
    =
    i f q_\mu \delta^{ab}\,. \label{eq:axial-current-overlap}
\end{align}
When the soft pion is extracted from an axial-current correlator, the coefficient of the one-pion pole is fixed by \cref{eq:axial-current-overlap}. Since $f$ is not renormalized, this coefficient contributes no term to \cref{eq:pion-wilson-coefficient-running}.

\section{Flavor-dressed double-soft pion theorem}\label[appendix]{sec:appendix-full-double-soft-pions}
This appendix records the full flavor-dressed form of the loop level corrections to double-soft pion theorem in the Chiral EFT. The double-soft theorem can be expressed as a soft operator acting on each external leg in the hard amplitude \cite{Weinberg:1966gjf,Cachazo:2015ksa,Rodina:2021isd,Berean-Dutcher:2025ohp}
\begin{align}
    \lim_{\tau\rightarrow 0}\cA^{a_1 a_2\cdots a_n a b}_{n+2}(\tau)= \sum_{i=1}^n\left(\NLSMSoftLeading + \tau \NLSMSoftSubleading\right)^{ab a_i d} \cA^{a_1\cdots d\cdots a_n}_{n}\,, \label{eq:double-soft-theorem-flavor-dressed-nlsm-rg}
\end{align}
where $a$ and $b$ label the flavor indices of the soft pions with momenta $\tau q_a$ and $\tau q_b$, respectively.

The one loop corrections enter at the same power counting as four-derivative operators in the Chiral EFT through a factor
\begin{align}
   \NLSMSoftSubleadingFlavor{ab a_i d}\supset \frac{1}{f^4}\frac{(p_i\cdot q_a)(p_i\cdot q_b)}{2p_i \cdot q_{ab}}\mathcal F_{(4)}^{ab a_i d}\,,
\end{align}
where $a_i$ is the flavor index of the hard leg with momentum $p_i$. At tree level, this correction was first obtained in \cite{Rodina:2021isd}. Introducing the shorthand notation
\begin{align}
    \log_{(ai)}&=\frac{1}{(4\pi)^2}\log\left(\frac{-\mu^2}{2p_i\cdot q_a\tau}\right),\nn\\
    \log_{(bi)}&=\frac{1}{(4\pi)^2}\log\left(\frac{-\mu^2}{2p_i\cdot q_b\tau}\right),
\end{align}
we can write the loop level corrections as
\begin{align}
    &\mathcal F_{(4)}^{ab a_i d}=\nn\\
    &\quad(\delta^{aa_i}\delta^{bd}+\delta^{ad}\delta^{ba_i})\left(\alpha_1+ \log_{(ai)}+\log_{(bi)}\right)\nn\\
    &+\delta^{ab}\delta^{a_i d}\left(\alpha_2+ \log_{(ai)}+\log_{(bi)}\right)\nn\\
    &+\big(T^{a a_i b d}+T^{a d b a_i}\big)\left(\alpha_3+ \frac{N}{6}(\log_{(ai)}+\log_{(bi)})\right)\nn\\
    &+\big(T^{a b d a_i}+T^{a a_i d b}\big)\left(\alpha_4+ \frac{N}{3}\log_{(ai)}\right)\nn\\
    &+\big(T^{a b a_i d}+T^{a d a_i b}\big)\left(\alpha_4+ \frac{N}{3}\log_{(bi)}\right).
\end{align}
Here $T^{abcd}=\langle t^at^bt^ct^d\rangle$, and the coefficients $\alpha_i(\mu)$ include the scheme-dependent local contributions.
In $\overline{\rm MS}$, using operator basis defined in \cref{eq:chiral-lagrangian-in-text}, they are
\begin{align}
\alpha_1(\mu)&=16\left(2L_1(\mu)+L_2(\mu)\right)+\frac{1}{4\pi^2}\nn\\
    \alpha_2(\mu)&=32L_2(\mu)+\frac{1}{4\pi^2}\nn\\
    \alpha_3(\mu)&=16L_3(\mu)+\frac{N}{(4\pi)^2}\frac{5}{9}\nn\\
    \alpha_4(\mu)&=8\left(L_3(\mu)+2L_4(\mu)\right)+\frac{N}{(4\pi)^2}\frac{13}{18}\,,
\end{align}
where $L_i$ are now the renormalized Wilson coefficients. We verified the double-soft theorem explicitly for the six pion amplitude at one loop order. 

From this full flavor-dressed soft theorem one can readily obtain the corresponding theorem for flavor-ordered amplitudes. 
The main text focuses on the planar single-trace case with two adjacent soft legs, for which the general expression above reduces to the simpler structure discussed in \cref{sec:chiraleft}.
\makeatletter
\interlinepenalty=10000
\bibliography{biblio}
\makeatother
\end{document}